\documentclass[aps,prx,reprint,superscriptaddress]{revtex4-2}

\usepackage{graphicx}
\usepackage{dcolumn}
\usepackage{bm}
\usepackage{amsmath}
\usepackage{hyperref}
\usepackage{xcolor}
\hypersetup{colorlinks,breaklinks,linkcolor=blue,urlcolor=blue,anchorcolor=blue,citecolor=blue}

\newcommand{\br}{\mathbf{r}}

\newcommand{\bk}{\mathbf{k}}

\begin{document}

\title{A Free Energy Approach to Modelling Disordered Hyperuniform Patterns}

\author{Alex Mamaev}
 \affiliation{%
 Institute of Scientific Computing, TU Dresden, 01062 Dresden, Germany
}
 \author{Enno Igel}
\affiliation{%
 Institute of Scientific Computing, TU Dresden, 01062 Dresden, Germany
}
 \author{Axel Voigt}%
 \affiliation{%
 Institute of Scientific Computing, TU Dresden, 01062 Dresden, Germany
}
 \author{Marco Salvalaglio}%
 \affiliation{%
 Institute of Scientific Computing, TU Dresden, 01062 Dresden, Germany
}
 \affiliation{Dresden Center for Computational Materials Science, TU Dresden, 01062 Dresden, Germany}

\date{\today}


\begin{abstract}
    We present a free-energy-based approach for modelling disordered hyperuniform patterns. Specifically, we devise a phase-field crystal model with a correlation kernel that simultaneously selects for a continuous range of degenerate wave modes. At small wave vectors, we construct the correlation function by inverting the Ornstein-Zernike relation to approximate the structure factor of a hyperuniform pattern. Our model demonstrates a crossover between liquid, crystalline, quasicrystalline, and long-lived disordered states depending on the specified free energy parameter and driving force. We show that, for a fixed driving force, the specific structures and wave vectors selected by the density field can be controlled by modulating the width of the correlation kernel. Furthermore, we show that the scaling of the low-wave-vector region of the correlation function can accurately reproduce a specified hyperuniform exponent, independently of the structural properties at larger wave numbers. Our results establish a free-energy-based framework for describing disordered hyperuniform states, demonstrating and enabling independent control over structural selection and long-wavelength density correlations.
\end{abstract}

\maketitle

\section{Introduction}\label{sec_1_Introduction}

Amorphous structures retaining correlations are ubiquitous in physical systems, yet their structural characterization remains a fundamental challenge \cite{debenedetti_supercooled_2001,berthier_theoretical_2011}.
A particularly intriguing class of such structures can be classified through the notion of hyperuniformity, a form of hidden order characterized by the suppression of density fluctuations at large length scales \cite{torquato_local_2003,torquato_hyperuniform_2018}.
While hyperuniformity is a hallmark of crystals and quasicrystals, it is generally absent in conventional disordered systems, such as liquids and gases.
Remarkably, hyperuniformity can also emerge in amorphous solids, including amorphous silicon and certain metallic glasses \cite{chen_disordered_2023}.
The coexistence of structural disorder and suppressed long-wavelength density fluctuations makes these materials particularly appealing, as it can enable unusual photonic, phononic, electronic, thermal, and transport properties while retaining statistical isotropy \cite{noauthor_isotropic_nodate,yu_engineered_2021,zhang_approach_2023,chen_disordered_2023,Phonons2017}.
Tuning hyperuniformity has also been exploited towards top-down engineering of materials with tailored physical properties, as demonstrated, for instance, for optical applications \cite{piechulla2021tailored,yu2021engineered,Vynck2023}.

In condensed matter systems, amorphous structures often emerge as metastable states through the kinetic arrest of positional or topological disorder, as exemplified by rapid quenching in glasses. Similarly, disordered hyperuniform (HU) configurations can arise through various nonequilibrium processes, including jamming, kinetic arrest, glassy dynamics, stationary active regimes, and defect-mediated transformations \cite{marcotte_nonequilibrium_2013,Nizam_2021,chen_topological_2021,kim_effect_2018,torquato_hyperuniform_2018,Gonzalo_HU_2019,backofen2024nonequilibrium}. Furthermore, suitably designed interactions can stabilize disordered HU glassy states with energies lower than those of competing crystalline or quasicrystalline configurations \cite{zhang_perfect_2016}. These observations highlight that hyperuniformity, as a property of large-scale density fluctuations, neither uniquely determines local structural arrangements \cite{salvalaglio_persistent_2024,milor_topological_2025} nor prescribes specific formation pathways or thermodynamic properties.

For regular amorphous solids, many open questions remain, including how these states undergo phase transitions, what structural signatures and defects they possess, and how to describe them using local information such as atomic potentials or liquid-state theory. This is particularly true for disordered HU systems, where structural information is even more ambiguous, and the metrics associated with hyperuniformity might shift during kinetic processes and as materials relax. 

Hyperuniformity has been detected in several experimental systems \cite{torquato_hyperuniform_2018}. However, systematically generating and experimentally characterizing disordered HU structures, particularly across phase transitions, remains challenging. As a result, a variety of numerical modeling techniques have been proposed to study HU materials computationally. For example, collective-coordinate methods generate disordered HU structures by directly constraining selected Fourier components of the density and can produce configurations with prescribed structure-factor scaling \cite{uche_constraints_2004,uche_collective_2006,aaron_fast_2024}. Such methods are well suited for generating specific HU clouds of points and of objects with prescribed shapes with minor modifications \cite{kim2020multifunctional,kim_ultradense_2025}. HU fields can be obtained by suitably designed Gaussian random fields \cite{ma_random_2017,milor_topological_2025}. These approaches, however, are not generally formulated to describe phase transformations or their time-dependent microstructural evolution and thus are not suited to relate HU fields to such regimes and physical properties.

Field theories have been widely used to study pattern formation across a broad range of systems \cite{cross_pattern_1993}, including ordered structures described by the Swift--Hohenberg model and its extensions \cite{swift_hydrodynamic_1977}, reaction--diffusion systems exhibiting Turing patterns \cite{turing_chemical_1952}, and phase-separating systems undergoing spinodal decomposition \cite{cahn_free_1958,cahn_spinodal_1961}. In these approaches, spatial structures emerge and evolve through the dynamics of one or more continuous fields, with equilibrium and relaxational models often formulated in terms of an underlying free-energy functional \cite{hohenberg_theory_1977}. The correlations retained in these theories render the emerging patterns classic examples of HU systems \cite{ma_random_2017}. In this context, the phase-field crystal (PFC) model \cite{elder_modeling_2002,elder_modeling_2004,elder_phase-field_2007} (or conserved Swift-Hohenberg model) is particularly well suited for studying microstructural evolution on atomic length scales and diffusive timescales. PFC models naturally describe a rich phenomenology, including order-disorder transitions, as well as elasticity and plasticity, for systems with order emerging from suitably defined non-local interactions in the free energy. Despite its relatively simple formulation, the PFC model can capture both the formation and subsequent evolution of structurally disordered phases, including structural glasses \cite{berry_simulation_2008,berry_modeling_2011,berry_phase-field-crystal_2014}, as well as other generic amorphous configurations and their temporal evolution \cite{toth_polymorphism_2010,abdalla_thermodynamics_2022}. Recently, HU scaling was demonstrated in a PFC model where structural frustration was introduced through a second atomic species acting as a disordered substrate \cite{mkhonta_liquid-substrate_2024}. These works suggest that PFC models can produce disordered HU configurations, although the relationship between the prescribed free-energy kernel, the resulting local structure, and the low-$k$ scaling remains largely unexplored.

In this paper, we employ a novel PFC model that can generate disordered HU patterns in a single-component system. This model enables not only the generation of a wide variety of patterns with controllable HU characteristics, but also the study of how they change phase while examining classical condensed-matter regimes such as solidification and growth. The paper is organized as follows: In Sec.~(\ref{sec_2_HU}), we introduce the formal concept of hyperuniformity, including how we will measure it in the context of PFC simulations; in Sec.~(\ref{sec_3_PFC}), we introduce the PFC model we employ; in Sec.~(\ref{sec_4_results}), we demonstrate how our PFC model can produce long-lived amorphous configurations whose hyperuniformity may be controlled through the model parameters independently of the structure. Finally, we conclude by discussing the main findings and provide some perspective on future directions in Sec.~(\ref{sec_5_Conclusion}).

\section{Hyperuniformity}
\label{sec_2_HU}

For a given cloud of $N$ points in real space with coordinates $\mathbf{r}_j$, hyperuniformity is characterized by the scaling of the number variance $\sigma_N^2(R)$ relative to the volume of an observation window of size $R$. Specifically, for a system in $d$ dimensions, $\lim_{R\to\infty}\sigma_N^2(R)/{R^d}=0$. An equivalent reciprocal-space characterization is provided through the structure factor, $S(\mathbf{k})=\left<\hat{\rho}(\mathbf{k})\hat{\rho}(-\mathbf{k})\right> /N$ where
\begin{equation}
    \hat{\rho}(\mathbf{k})=\sum_{j=1}^{N}e^{-i\mathbf{k}\cdot\mathbf{r}_j},
\end{equation}
is the Fourier transform of the microscopic density $\rho(\mathbf{r})=\sum_j^N \delta(\mathbf{r}-\mathbf{r}_j)$, and $\bk$ is the wave vector (in the remainder of the paper we will follow the convention $k\equiv|\bk|$).
HU systems satisfy $\lim_{\mathbf{k}\to0}S(\mathbf{k})=0,$ and may be characterized through a power-law scaling at low $k$, $S(k)\sim k^\alpha$, where $\alpha$ is a characteristic exponent often used to classify different types of hyperuniformity in point patterns \cite{torquato_hyperuniform_2018}. 

While the above definition of hyperuniformity is formulated for point patterns, the concept can be naturally extended to continuous field representations of the density \cite{ma_random_2017,torquato_hyperuniform_2018}. In our case, the discrete positions $\mathbf{r}_j$ will be replaced by a coarse-grained density field with maxima in such positions. The precise structure of how different peaks are interpolated is typically encoded in Fourier modes larger than $2\pi/a$, where $a$ is the lattice constant (or an effective atomic length scale for an amorphous system). As a result, the low-wavenumber spectral behavior is still expected to be recovered when we consider some coarse-grained version of the density field,
\begin{equation}\label{eq_smoothed_density}
    n(\br) = \int d\bk e^{i\bk\cdot \br} \hat{\rho}(\bk)e^{-k^2/2\sigma^2}
\end{equation}
with smoothing width $\sigma$. If we analyze the spectral density of such a field, $\hat{\chi}(\mathbf{k}) = \left<\delta\hat n(\mathbf{k})\delta\hat n(-\mathbf{k})\right>$/V, where $\delta\hat n(\mathbf{k})$ is the Fourier transform of $ n(\mathbf{r})-\left<n(\mathbf{r})\right>$ and $V$ is the volume of the system. Hyperuniformity can be characterized by $\lim_{\bk\to0}\hat{\chi}(\mathbf{k})=0$, or by the exponent $\alpha$ for a spectral density which scales as $\hat{\chi}(k)\sim k^\alpha$ \cite{ma_random_2017,torquato_hyperuniform_2018}. 

In the remainder of this paper, when we analyze PFC density fields, we assume they approximate a coarse-grained atomic density akin to Eq.~\eqref{eq_smoothed_density}.

\section{Phase-Field Crystal (PFC) Model}\label{sec_3_PFC}
PFC models evolve a dimensionless, spatially-varying field $n\equiv n(\mathbf{r})$ akin to that defined in Eq.~\eqref{eq_smoothed_density} over diffusive timescales.
They may be related to the classical density functional theory of freezing \cite{elder_phase-field_2007} by deriving a free energy functional for fields $n=(d-d_0)/d_0$ via a perturbative approach about the liquid state having reference density $d_0$. A prescribed two-point direct density correlation function provides the minimal ingredient required for departures from a uniform state to a periodic state. 

For pure materials, a simple form of the PFC free energy functional can be written as follows \cite{xtal1,xtal2},
\begin{align} \label{eq_PFC_free_energy}
        \frac{\Delta F}{\rho_0k_B T} =\int d\mathbf{r}\bigg(&\frac{1}{2}n^2-\frac{1}{6}n^3+\frac{1}{12}n^4\nonumber\\
        &-\frac{1}{2}n\int d\mathbf{r'}C^{(2)}(\mathbf{r}-\mathbf{r'})n\bigg).
\end{align}
Here, the cubic and quartic terms arise from a truncated expansion of a pure material density functional theory. The numerical prefactors for these terms are often treated as parameters to adjust the free energy landscape of the model, but here we hold them fixed to focus our analysis on the effects of the correlation function.

The correlation kernel $C^{(2)}(\mathbf{r}-\mathbf{r'})$ may be chosen to introduce interactions at a specific length scale and is more readily interpreted in Fourier space. 
For example, simple polynomial kernels in Fourier space have been shown to possess the necessary ingredients to reproduce the expected behaviors relevant to the thermodynamics of solid-liquid phase transitions and a number of solid-state phenomena, including dendritic solidification, coarsening, elasticity, plasticity, among others \cite{elder_modeling_2002,elder_modeling_2004,elder_phase-field_2007,skogvoll_phase_2022}

Interestingly, when the form of the kernel is chosen to have the simplest possible form consisting of a single peak in Fourier space, i.e., $\hat{C}^{(2)} (k)= C_0+C_2 k^2-C_4 k^4$, the PFC model has demonstrated the capacity to produce long-lived amorphous states, particularly in three spatial dimensions \cite{abdalla_thermodynamics_2022}. However, the referenced work showed that the amorphous structures depended on a large-amplitude noise parameter used to initialize the system. In other works involving higher-order correlation kernels (multiple length scales), amorphous cluster formation has also been observed, although such states eventually coarsen into BCC/FCC phases \cite{toth_polymorphism_2010,toth_amorphous_2011}. 

While these polynomial kernels are purely phenomenological, they can be connected to more realistic correlation kernels or used to reproduce desired structural effects within the model. In traditional liquid-state theory, the structure factor and direct correlation function are related through the Ornstein-Zernike relation \cite{hansen_theory_2013},
\begin{equation}\label{eq_OZ}
    S(k)=\frac{1}{1-\hat C^{(2)}(k)}.
\end{equation}
This relation has been shown to hold accurately for soft-core interactions \cite{Mkhonta2016}, whereas it remains an approximation for other types of interactions. $\hat{C}^{(2)}(k)$ incorporates the reference density ($d_0$), ensuring a dimensionless free-energy functional, as is commonly used in PFC models. Thus, the scaling of the PFC correlation kernel at small wave-vector $k$ can be engineered to control the hyperuniformity exponent, such that $S(k)\sim k^\alpha$. 
Specifically, one can choose a prescribed exponent $\alpha$ for the structure factor at low $k$ and invert it through Eq.~\eqref{eq_OZ}, choosing a PFC correlation function that scales as $1-k^{-\alpha}$ at low $k$. This, in principle requires a divergent $\hat{C}^{(2)} (k)$ at $k=0$. To remedy this, we employ a cutoff, defined below a specific $k$, where our correlation takes on a constant negative value.

\begin{figure}[h]
    \centering
    \includegraphics[width=0.9\linewidth]{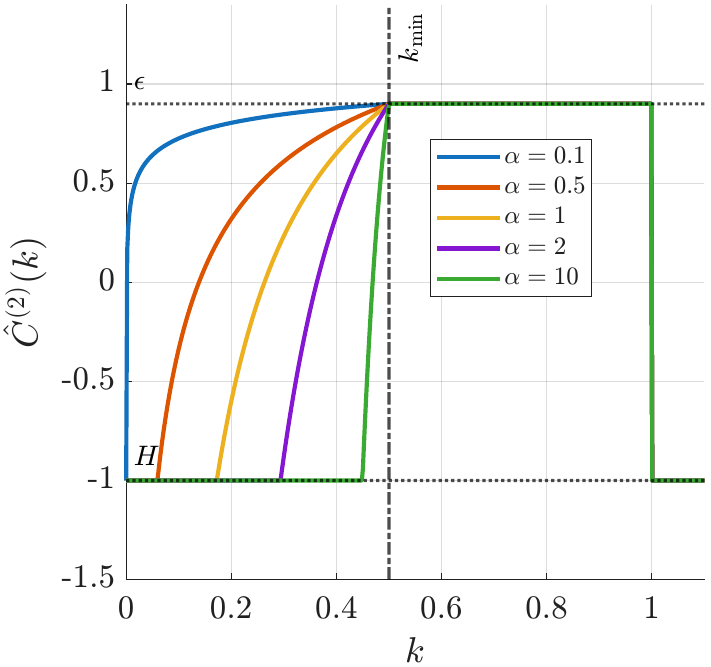}
    \caption{Here we show the two-point density-density correlation function $\hat{C}^{(2)}(k)$ as a function of wave-vector $k$, described in Eq.~\eqref{eq_plateau_kernel}. For demonstrative purposes, we employ a value of $H=-1$, $k_{\min}=0.5$, $\epsilon=0.9$, $A=1$, and 5 choices of $\alpha$ ranging from $0.1$ to $10$  }
    \label{fig_correlations}
\end{figure}

The additional ingredient required for our model is a method to produce long-lived disordered structures, i.e., enforcing some degree of correlated disorder at small to intermediate scales. To do this, we use a similar choice for $S(k)$ as in \cite{salvalaglio_persistent_2024}. Specifically, we consider a plateau in the correlation such that $\hat{C}^{(2)} (k)=\epsilon$ for all $k_{\min}\leq k < 1$. The resulting correlation function of our model is
\begin{equation}
\label{eq_plateau_kernel}
\hat{C}^{(2)}(k) = 
\begin{cases} 
\max \left[ H, \Phi(k) \right] & k \leq k_{\min} \\
\epsilon & k_{\min} < k < 1 \\
H & k \geq 1
\end{cases},
\end{equation}
where $\Phi(k) = \epsilon - A \left[ (k_{\min}/k)^\alpha - 1 \right]$ and $A$ a parameter. We plot the correlation function for varying choices of $\alpha$ in Fig.~\ref{fig_correlations}. 
The corresponding structure factor expected from this correlation kernel, within the unregularized low-$k$ interval over which $\hat C^{(2)}(k)=\Phi(k)$ is,
\begin{equation}
\label{eq_plateau_structure_factor}
    S(k)=\frac{1}{1-\epsilon + A \left[ (k_{\min}/k)^\alpha - 1 \right]}.
\end{equation}

In this parametrization, $\epsilon$ acts as an effective temperature or driving force for pattern formation. The parameter $H$ sets the regularized value of the correlation function at sufficiently small wave number and imposes a strong penalty on modes with $k\geq1$. Together with $H$ and $\alpha$, $A$ controls the wave number at which the low-$k$ correlation function reaches the cutoff. Finally, $k_{\min}$ defines the lower boundary of the plateau of energetically preferred wave numbers.

For $k_{\min}$ sufficiently close to 1, this model is expected to behave similarly to a conventional single-mode PFC model and form periodic crystalline states with a single length-scale. Decreasing $k_{\min}$ progressively broadens the spectrum of accessible long-wavelength relaxation modes, thereby frustrating conventional ordering pathways and promoting coexistence between competing local structures. As will be shown in Sec.~(\ref{sec_4_results_varying_kmin}), this produces a sequence of increasingly frustrated states, including crystalline and amorphous configurations.

We also note that Eq.~\eqref{eq_plateau_structure_factor} is approximate. Away from the homogeneous reference state, we expect the nonlinearities in Eq.~\eqref{eq_PFC_free_energy} to prevent the measured spectral density from being identified directly with the liquid-state theory expression, as discussed further in Sec.~(\ref{sec_4_results_varying_alpha}).

\subsection{Dynamics}

The density field is evolved according to conserved stochastic gradient-flow dynamics \cite{elder_modeling_2002},
\begin{equation}\label{eq_PFC_diffusive_dynamics}
    \frac{\partial n}{\partial t}
    =M\nabla^2\frac{\delta \mathcal{F}}{\delta n}+\eta,
\end{equation}
where $\mathcal{F}=\Delta F/(\rho_0k_BT)$ is the dimensionless free energy, $M$ is a spatially constant mobility, and $\delta\mathcal{F}/\delta n$ is the corresponding chemical potential. The conserved stochastic noise $\eta(\mathbf{r},t)$ satisfies
\begin{align}
    \langle\eta(\mathbf{r},t)\rangle &=0, \nonumber\\
    \langle\eta(\mathbf{r},t)\eta(\mathbf{r}',t')\rangle
    &=-2M\Theta\nabla^2
    \delta(\mathbf{r}-\mathbf{r}')\delta(t-t'),
\end{align}
where $\Theta$ is the dimensionless noise strength. Without loss of generality, in the simulations we choose $\Theta\propto 1-\epsilon$, such that the fluctuation amplitude decreases as the effective driving force for pattern formation increases.

The inclusion of stochastic fluctuations is particularly important here, as the resulting amorphous states are strongly influenced by kinetic pathway selection, metastability, and nonlinear mode competition. Small fluctuations may therefore significantly alter the resulting local ordering pathways and the final metastable configurations obtained during evolution, in agreement with recent PFC results on amorphous patterns \cite{abdalla_thermodynamics_2022}.

\section{Results and Discussion\label{sec_4_results}}

In this section, we present our simulation results, focusing on four key aspects of the model and discussing the corresponding results. First, we characterize the ordered and disordered configurations stabilized across the parameter space defined by $n_0$ and $\epsilon$, demonstrating the versatility of the proposed minimal model. Second, we demonstrate that the hyperuniformity exponent $\alpha$, extracted from the spectral density, can be largely tuned and matches the value prescribed by the correlation function in Eq.~\eqref{eq_plateau_kernel}. Third, we analyze the structural features encoded in the spectral density over the intermediate wave-vector range $k_{\min} < k < 1$ to gain insight on the emerging patterns. Finally, we investigate how varying the temperature parameter $\epsilon$ controls solidification, enabling a transition from ordered crystalline structures to disordered configurations, serving as proof of concept for energy driven material processes.

All simulations were performed in two spatial dimensions on a square periodic domain of size $4096\times4096$. The density field was discretized using a uniform Cartesian mesh with spatial discretization $\Delta x = 2\pi/8$
corresponding to approximately eight grid points per dominant lattice spacing associated with the upper unstable wave number cutoff at $k=1$. Unless otherwise stated, we fix the correlation function parameters to $\alpha=2$, $H=-1\times10^{12}$, $A=1$.

We solved Eq.~\eqref{eq_PFC_diffusive_dynamics} using a semi-implicit Fourier spectral scheme with a fixed timestep of $\Delta t = 0.1$ throughout all simulations. Unless otherwise specified, we report below patterns obtained by relaxing the initial conditions via Eq.~\eqref{eq_PFC_diffusive_dynamics} for $10^5$ timesteps. At this point, variations in the system's free energy, averaged over a number of subsequent timesteps, fall below a specified tolerance consistent with effective stationary patterns.

\subsection{Structural diversity under varying driving force}

In traditional PFC models, the parameters $n_0$ and $1-\epsilon$, analogous to an average density and dimensionless temperature, provide the driving force for solidification/crystallization. By constructing an explicit ansatz for the density field, typically a truncated Fourier series whose wave modes prescribe the dominant length scales, one can insert it into the free energy and, upon coarse-graining, construct an analytical approximation to the system's phase diagram. With the correlation kernel employed in this work, however, all wave-vectors in the range $[k_{\min},1]$ are degenerate. Furthermore, any amorphous configuration does not permit a simple real-space or Fourier-space description of a structural element by definition. As a result, we numerically characterize the spectrum of metastable configurations realized across the parameter range we consider. 

\begin{figure*}
    \centering
    \includegraphics[width=0.975\linewidth]{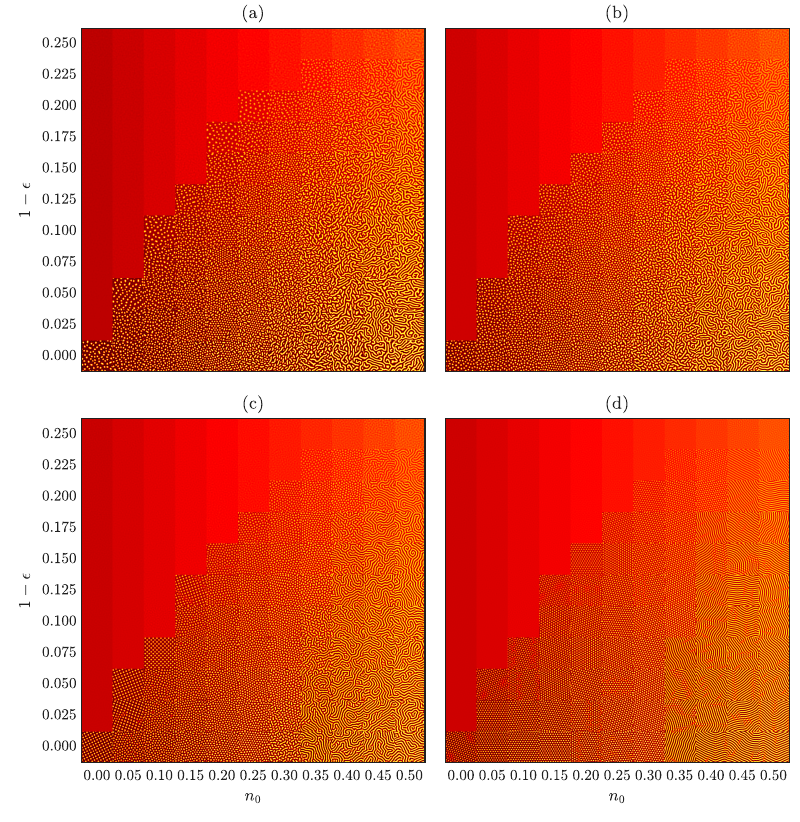}
    \caption{Representative density configurations after an initially liquid state relaxes into its long-lived configuration for different choices of $n_0$ (x-axis) and $\epsilon$ (y-axis) with (a) $k_{\min}$ = 0.4, (b) $k_{\min}$ = 0.55, (c) $k_{\min}$ = 0.65, and (d) $k_{\min}=0.9$.}
    \label{fig_phase_diagram}
\end{figure*}

Fig.~\ref{fig_phase_diagram} shows the long-lived density configurations for different choices of $n_0, \epsilon$, and four choices of $k_{\min}$ between $k_{\min}=0.4$ and $k_{\min}=0.9$. We first observe that the model recovers the transition between the liquid phase, associated with a nearly constant $n$, and the solid phase, where significant density oscillations and peaks are observed. Various degrees of order are obtained. For narrow choices of our kernel, we observe the traditional liquid, hexagonal, and stripe phase regions expected from a single-peak PFC model. However, when we allow a wider correlation kernel, we observe a much more diverse landscape of structures accessible to the model, including liquid, amorphous, hexagonal, square, dimer, quasicrystalline, and stripe phases commonly associated with multiple length scales \cite{mkhonta_exploring_2013}. 
At large $n_0$, patterns akin to bicontinuous and spinodal structures can also be observed. 

\begin{figure*}
    \centering
    \includegraphics[width=\linewidth]{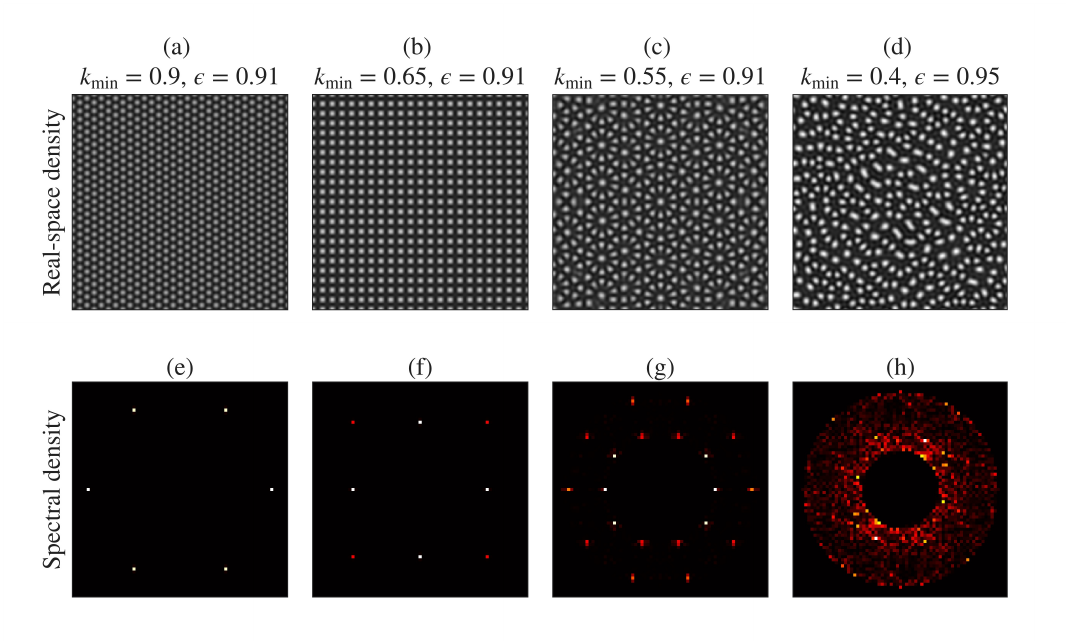}
    \caption{Example of ordered and correlated-disordered structures obtained as equilibrium / metastable states. Four cases are reported with real-space density configurations shown from panel (a) to (d), and the corresponding spectral densities in the panels immediately below, panels (e) to (h). The reported configurations are obtained by initializing a seed crystal that is allowed to grow to fill the entire simulation box. The initialized wave vectors chosen for the mode expansions correspond to the first and second dominant modes for (a) a hexagonal phase, (b) a square phase, and (c) a decagonal quasicrystalline phase. The amorphous structure in panel (d) is initialized with noise. In panels (e) to (h), we plot the domain $k_x\in [-1,1]$, $k_y\in [-1,1]$, where $k_x$ and $k_y$ are the $x$ and $y$ components of the wave-vector $k$, respectively. The chosen $k_{\rm min}$ and $\epsilon$ are reported in the panels.}
    \label{fig_structure_gallery}
\end{figure*}

The stabilization of distinct crystalline, quasicrystalline, and disordered HU arrangements is further illustrated in Fig.~\ref{fig_structure_gallery}. We initialize the density field to reproduce a small seed with given symmetry/correlations using mode expansions corresponding to different crystalline and quasicrystalline symmetries, varying $k_{\rm min}$ and $\epsilon$ (for constant $n_0=0.05$). Depending on the temperature and the choice of $k$, these configurations either grow towards stable triangular, square, or tenfold quasicrystalline structures, or maintain a disordered-correlated (HU) character. We thus also identify stretched HU configurations that persist as stationary states and result from growth (see further discussion in Sec.~(\ref{sec_4_results_solidification})). All reported stationary configurations are converged to numerical precision.

\subsection{Control of hyperuniform scaling exponent $\alpha$}\label{sec_4_results_varying_alpha}
Having demonstrated the ordered and disordered configurations stabilized by the correlation kernel of Eq.~\eqref{eq_plateau_kernel}, we now demonstrate effective control over the scaling of the HU exponent of the spectral density by varying $\alpha$.

\begin{figure*}
    \centering
    \includegraphics[width=\linewidth]{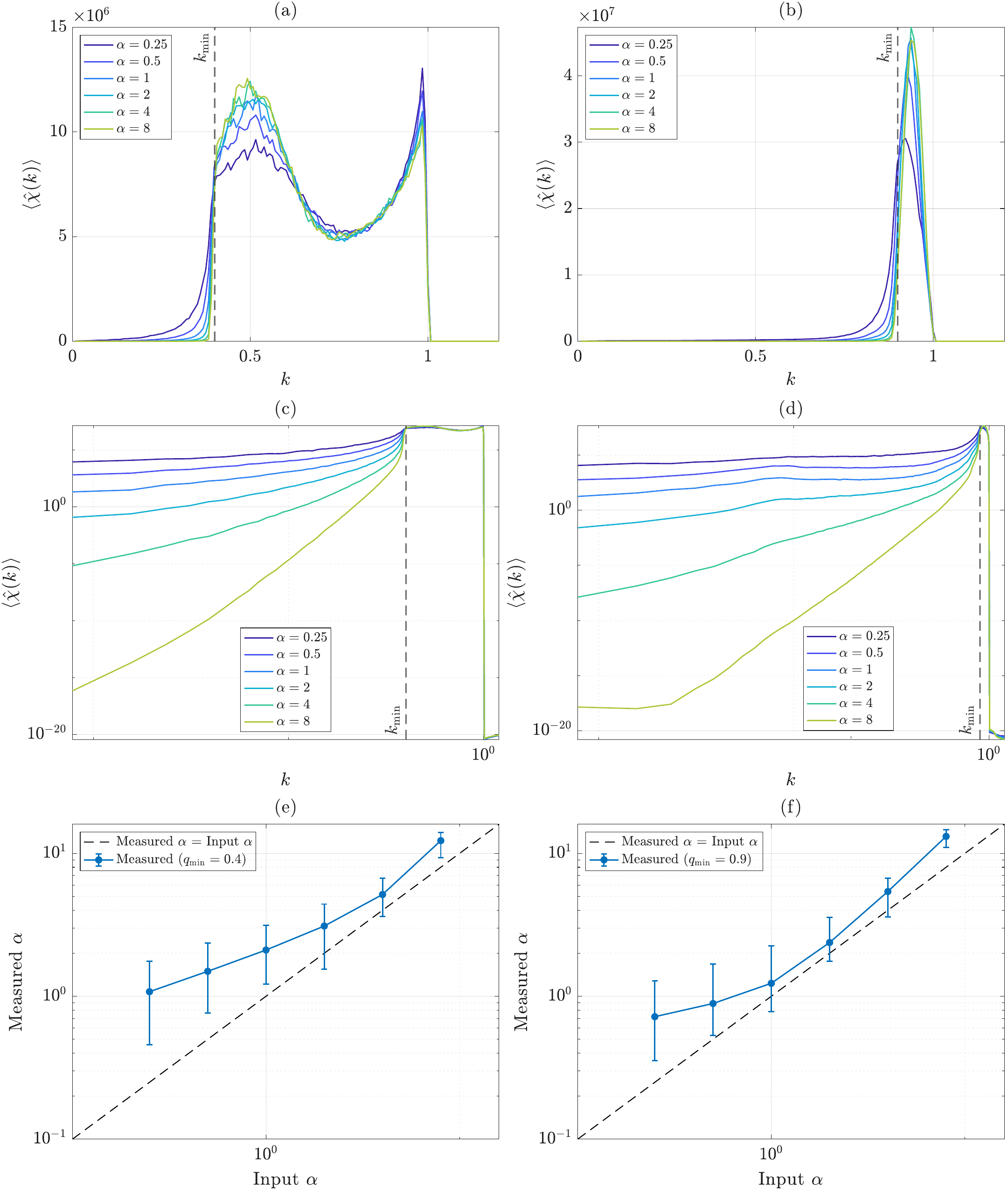}
    \caption{Spectral densities $\hat{\chi}(k)$ from PFC simulations with different choices of the input HU exponent $\alpha$ for (a) $k_{\min}=0.4$ and (b) $k_{\min}=0.9$. Also shown are the HU exponents $\alpha$ extracted from the PFC simulations by linearly fitting the resulting spectra between $k_c$ and $k_{\min}$. The extracted exponents are plotted against the introduced exponent in Eq.~\eqref{eq_plateau_kernel} for (c) $k_{\min}=0.4$ and (d) $k_{\min}=0.9$. }
    \label{fig_alpha_vs_alpha}
\end{figure*}

We initialize our system with different values of $\alpha$, averaging over 20 realizations of Gaussian uncorrelated noise as the initial condition. We consider a fixed point ($n_0,\epsilon$)=(0.05,0.95) close to the solid/liquid coexistence region for two distinct intervals of favored wavelengths, namely $k_{\rm min}=0.4$ and $k_{\rm min}=0.9$, see Fig.~\ref{fig_phase_diagram}(a) and Fig.~\ref{fig_phase_diagram}(d), respectively. In Fig.~\ref{fig_alpha_vs_alpha}, we show the dependence of the exponent achieved by the spectral density $\hat{\chi}(k)$ on the $\alpha$ chosen in our correlation function. The exponent is calculated by a linear fit in log-log space, where the fitting window range is determined as follows.

The lower bound is chosen as a function of the exponent $\alpha$, as this exponent influences the wave-vector at which the correlation function is cut off and effectively becomes flat,
\begin{equation}\label{eq_wave_vector_cutoff}
    k_{\text{c}} = k_{\min}(1+(\epsilon-H)/A)^{-1/\alpha}.
\end{equation}
For finite $H$, the regularization produces a small-$k$ plateau $S(k\to0)=(1-H)^{-1}$. Hence, the model is not ideally HU in the infinite-system limit. Instead, it realizes effective HU scaling over the finite interval $k_c<k<k_{\min}$. The magnitude of the residual long-wavelength fluctuations is controlled by $H$, while the width of the scaling regime is controlled jointly by $H$, $A$, and $\alpha$. For sufficiently large $\alpha$, the cutoff enters the numerically resolved spectrum and reduces the available interval over which the prescribed power-law scaling can be observed. We treat the upper bound as a variable quantity, from the lowest upper bound set to 3 discrete Fourier grid modes above $k_c$ to the highest upper bound set equal to $k_{\min}$. The error bars in Fig.~\ref{fig_alpha_vs_alpha} represent the range of $\alpha$ extracted from our PFC simulations due to variations in these fitting windows, along with the statistical error from the fit.  

As a result of the strong, nonlinear scaling of $k_c$ with $\alpha$, for progressively larger $\alpha$, $k_c$ asymptotically approaches $k_{\min}$. In this limit, the power-law region becomes substantially narrower, and the nonlinearities of the free energy of Eq.~\eqref{eq_PFC_free_energy} become more pronounced than under the assumed linear behavior in the low-$k$ limit. Thus we observe a systematic overestimation of the slope for sufficiently wide choices of $\alpha$. 

The results in Fig.~\ref{fig_alpha_vs_alpha} indicate that our model captures the expected $\alpha$ scaling within the parameter range we consider. Furthermore, we remark that such scaling can be controlled regardless of the structure selected by the density configuration. Specifically, we show that the model reproduces the correct HU scaling for both ordered crystalline configurations ($k_{\min}=0.9$) and amorphous configurations ($k_{\min}=0.4$). This effective decoupling between long-range order and local arrangements allows exploration of HU pattern manifolds \cite{torquato_hyperuniform_2018}. It further underscores the multiplicity of local arrangements achievable for a given HU character and vice versa.
We also note that although $\alpha$ determines the low-wave-number scaling of the spectral density, variations in $\alpha$ also modify the shape of the correlation function near $k_{\min}$. These variations are more apparent at smaller values of $\alpha$, where the range of wave modes responsible for forming a particular structure may increase through a wider effective tail extending toward $k=0$. This behavior may be further controlled by an appropriate choice of the parameter $A$, which is fixed to a constant value in this work. In particular, increasing $A$ allows the slope of the correlation function at $k_{\min}$ to be made consistent across different choices of $\alpha$. This enables equivalent structural properties in the spectral density while allowing for parametric selection of the HU exponent $\alpha$.

\subsection{Structural property shifts through varying plateau width $k_{\min}$\label{sec_4_results_varying_kmin}}

The analysis reported in Fig.~\ref{fig_alpha_vs_alpha} shows that, depending on $k_{\min}$, various spectral densities are obtained featuring either one or two peaks; see in particular Fig.~\ref{fig_alpha_vs_alpha}(a,b). This behavior is further characterized and quantified in this section. We first show how the correlation function behaves as the frustration parameter $k_{\min}$ varies. We initialize the density field with a Gaussian random field with mean $n_0=0.05$ at $\epsilon=0.95$, close to the solid-liquid coexistence line, as in \cite{abdalla_thermodynamics_2022}, and evolve the system until it reaches effectively stationary patterns.

Overall, across the parameter range considered, the system exhibits a clear progression of ordering behavior. Specifically, for large $k_{\min}$ close to 1, we observe ordered crystalline systems (hexagonal and square phases). Below a particular threshold, we begin to observe long-lived amorphous configurations. These amorphous configurations are characterized by a wide range of wave modes that contribute to the resulting structure, with smaller variation in magnitude across them than in a crystalline state. 

\begin{figure*}
    \centering
    \includegraphics[width=\linewidth]{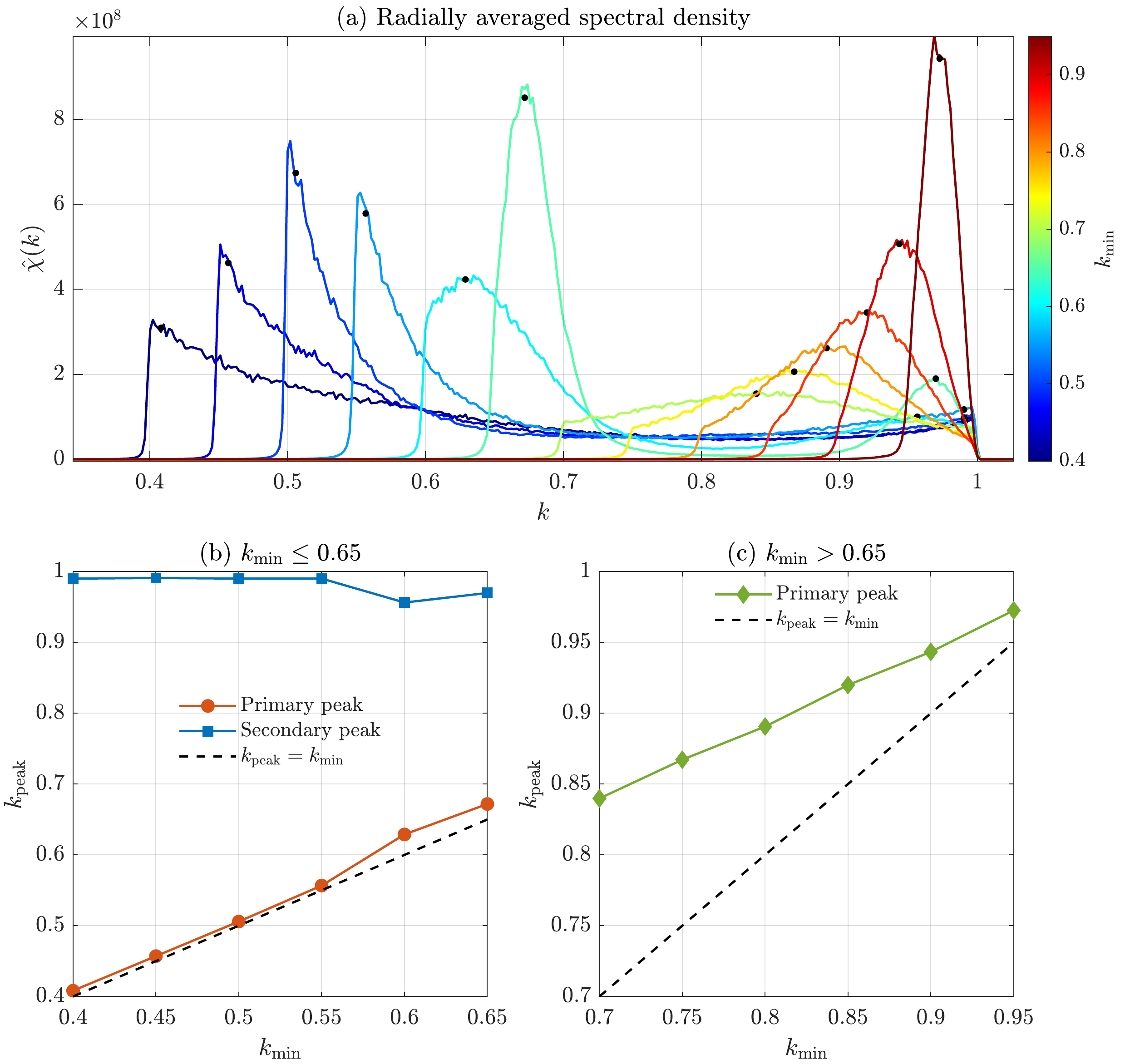}
    \caption{Structural properties of the emerging patterns. (a) radially averaged spectral densities $\hat{\chi}(k)=\langle n(\bk)n(-\bk)\rangle$ plotted against wave-vector $k$ for different choices of the plateau width parameter $k_{\min}$. Peak positions of the spectral density are then shown for (b) $k_{\min}$ below or equal to 0.65, where two prominent maxima exist, and (c) $k_{\min}$ above 0.65, where only a single prominent maximum exists.}
    \label{fig_structure_peak_ratio}
\end{figure*}

To quantitatively analyze the obtained structures, we show the resulting spectral densities in Fig.~\ref{fig_structure_peak_ratio}.
For narrower plateau choices, akin to traditional PFC models, the spectral density exhibits a single dominant structural peak located approximately at the midpoint of our correlation kernel's plateau ($k=(1+k_{\min})/2$). 
For wider plateaus, the density field selects a wide distribution of wave-vectors across the range $[k_{\min},1]$. We generally observe two prominent peaks which appear at $k=k_{\min}$ and $k=1$. Unlike traditional PFC models, these peaks do not correspond to ratios of the first and second Bragg peak wave-vectors in particular crystalline and quasicrystalline systems, where the nonlinear terms in Eq.~\eqref{eq_PFC_free_energy} provide additional energetic compensation beyond purely excess-energy contributions. Instead, we see a broad range of nonlinear couplings selected, leading to amorphous density-field configurations.

Here, the correlation kernel preferentially favors the formation of structures at $k_{\min}$, the least linearly stable mode. The second peak, located at $k=1$, may be attributed to the strong energetic penalty applied to wave vectors beyond $k=1$, leading to a pile-up of modes near this cutoff. Interestingly, the coexistence of two peaks at wave numbers corresponding to non-commensurate lengths closely resembles the correlations effectively imposed among the two species considered in Ref.~\cite{mkhonta_liquid-substrate_2024}, thus incorporating and generalizing such frustration of characteristic lengths. We note that the structural features described here may also depend on the remaining parameters of the correlation function. In particular, as discussed in Sec.~(\ref{sec_4_results_varying_alpha}), an appropriate choice of the parameter $A$, which we treat as fixed for most of this work, allows effectively independent control of the structural and HU properties of the model.

\subsection{Solidification properties\label{sec_4_results_solidification}}

Finally, we highlight the characteristic features of phase-formation kinetics emerging from our simulations. We consider the solidification properties of our model starting from an initial circular seed set to a hexagonal lattice with wave-number $k_{\min}$, with an initial radius of 5 lattice constants. We choose this wave-vector because it is the least-damped mode in a linear stability analysis with our kernel choice for $\epsilon<1$.

We consider different values of the temperature parameter $\epsilon$ for fixed $n_0$ to observe the role of the driving force on the solidification rate and the resulting structures. To quantify the structural changes, we calculate the global sixfold bond-orientational order parameter \cite{steinhardt_bond-orientational_1983}
\begin{equation}
    \Psi_6=\left|\frac{1}{N}\sum_{j=1}^{N}\frac{1}{N_j}\sum_{k=1}^{N_j}e^{6i\theta_{jk}}\right|,
\end{equation}
through a real-space analysis of the density field. Here, $N$ is the number of identified density peaks, $N_j$ is the number of nearest neighbors of peak $j$ (calculated via Delaunay triangulation), and $\theta_{jk}$ is the angle of the bond connecting peaks $j$ and $k$. We identify density peaks by thresholding the density field at half its maximum value, followed by a search for local maxima. 

\begin{figure}
    \includegraphics[width=\linewidth]{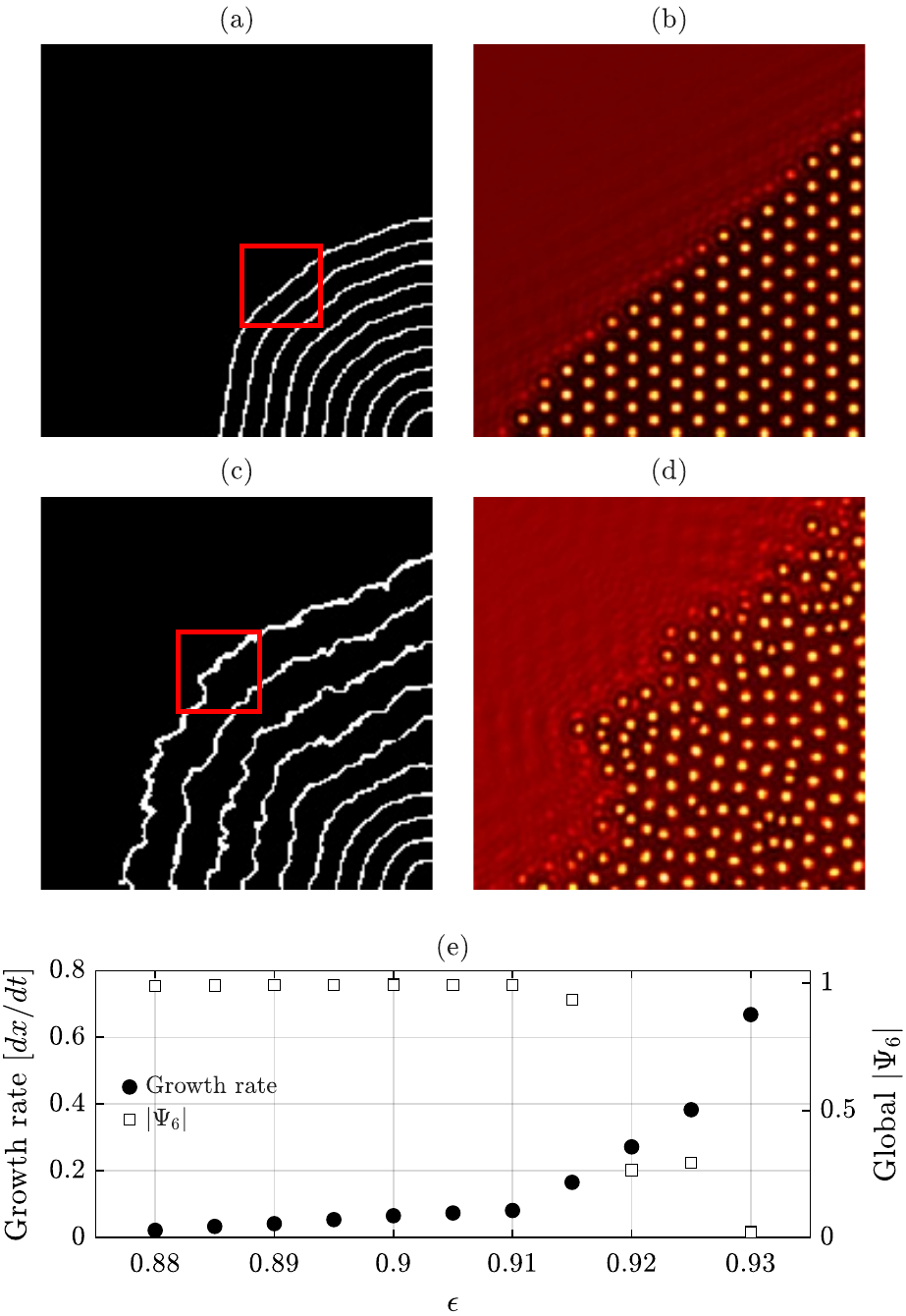}
    \caption{Growth profiles for equally spaced timesteps of one quarter of a solidifying system for (a) $\epsilon=0.91$ and (c) $\epsilon =0.915$. Zoomed-in versions of the density fields corresponding to the red boxed regions in panels (a) and (c) after 100000 timesteps are also shown in panels (b) and (d), respectively. The growth rate and six fold bond order parameter, $|\Psi_6|$ are also plotted as a function of $\epsilon$ in panel (e).}
    \label{fig_solidification}
\end{figure}

Our results are shown in Fig.~\ref{fig_solidification}. As $\epsilon$ increases up to $\epsilon = 0.91$, we observe steady growth of the initially seeded hexagonal configuration, with an approximately linear increase in growth rate with $\epsilon$ and a strong six-fold ordering. Beyond this $\epsilon$, we see a pronounced shift in the solidification rate and growth morphology: the seeded hexagonal configuration destabilizes and forms an amorphous structure, accompanied by a sharp decrease in global ordering, $\Psi_6$. This destabilization appears to be purely driven by nonlinearities, and the resulting spectral density of the solidifying amorphous system appears qualitatively similar to that of the amorphous phases emerging from initial fluctuations in Sec.~(\ref{sec_4_results_varying_kmin}). 
Notably, because the correlation function in Eq.~\eqref{eq_plateau_kernel} strongly suppresses large-wavelength interactions, solidification in the present model differs from traditional PFC models. This is because the average density, which is initialized as a uniform field, may not diffuse between the liquid and newly formed solid phases. Any such diffusion is strongly suppressed by the excess free-energy contribution; hence, there is no effective partitioning of the average density. As such, we do not compare with the scaling observed in marginal stability analysis \cite{galenko_marginal_2011}.

\section{Conclusion \label{sec_5_Conclusion}}

In this work, we have introduced a phenomenological framework for describing disordered hyperuniform phases, based on a plateau-shaped correlation function incorporated into a PFC-like free-energy functional. We demonstrated that the model supports transitions from a uniform state to either ordered or disordered patterned states, depending on the plateau width in the correlation kernel and the thermodynamic driving force. The resulting amorphous structures exhibit a broad, nearly continuous spectrum of contributing Fourier modes, with dominant spectral peaks near the edges of the plateau. Importantly, the proposed correlation function provides a minimal yet versatile description that can reproduce a wide variety of patterns, including many previously obtained within more specialized field-theoretical models, while offering independent control over long-wavelength scaling behavior.

Using the Ornstein-Zernike relation, we parametrized the low-$k$ region of our model's correlation function to match the scaling of the theoretical structure factor from liquid-state theory. We then showed that our model dynamically recovers the correct scaling exponent $\alpha$ for the resulting spectral density in the interval $k_c<k<k_{\min}$. Thus, our model provides a free-energy-based framework for generating and evolving structures with a prescribed effective HU scaling. Importantly, this scaling is independent of the value $k_{\min}$ chosen for the structural part of the correlation function, indicating that these two quantities, and thus local arrangement and long-wavelength density fluctuations, can be tuned independently. This suggests that the low-$k$ regions of traditional PFC correlation kernels and other energy-based field theories may also be modified using the same approach to reproduce a specific target HU exponent.

While the correlation function presented in Eq.~\eqref{eq_plateau_kernel} is strictly phenomenological, it demonstrates that length-scale frustration is a sufficient mechanism for generating amorphous structures with PFC. Based on the findings reported here, more specialized forms can then be considered, allowing the study of specific classes of materials while tuning specific material properties. It remains to be seen whether a continuous range of wave vectors is required to generate such structures with a free-energy-based approach or whether, through explicit off-resonance combinations of reciprocal-space peaks, one can construct more realistic correlation functions to generate amorphous density fields. More generally, the free-energy framework proposed here extends the PFC model toward a general description of disordered-correlated systems, providing a methodology for studying their growth, thermal relaxation, defects, and mechanical deformation, and how these affect local structure and low-wavenumber properties.

\section*{Acknowledgements}

We thank Simiso Mkhonta, Ken Elder, and Zhi-Feng Huang for useful discussions on PFC modeling of disordered structures and hyperuniformity. We gratefully acknowledge funding by the German Research Foundation (DFG) -- Project number 556185784, and computing time on the high-performance computer at the NHR Center of TU Dresden. This center is jointly supported by the Federal Ministry of Research, Technology and Space of Germany and the state governments participating in the NHR (\href{www.nhr-verein.de/unsere-partner}{www.nhr-verein.de/unsere-partner}).

\section*{Data Availability}

The data that support the findings of this work will be made openly available upon acceptance.

\bibliography{bibliography}

@article{milor_topological_2025,
      title={A topological approach to the Cahn-Hilliard equation and hyperuniform fields}, 
      author={Abel H. G. Milor and Otto Sumray and Heather A. Harrington and Axel Voigt and Marco Salvalaglio},
      year={2026},
      url = {https://doi.org/10.4171/ifb/580},
	doi = {10.4171/IFB/580},
	  journal={Interface and Free Bound., In press}
}

@article{Mkhonta2016,
  title = {Emergence of Chirality from Isotropic Interactions of Three Length Scales},
  author = {Mkhonta, S. K. and Elder, K. R. and Huang, Zhi-Feng},
  journal = {Phys. Rev. Lett.},
  volume = {116},
  issue = {20},
  pages = {205502},
  numpages = {5},
  year = {2016},
  month = {May},
  publisher = {American Physical Society},
  doi = {10.1103/PhysRevLett.116.205502}
}

@article{Phonons2017,
  title = {Hyperuniform disordered phononic structures},
  author = {Gkantzounis, G. and Amoah, T. and Florescu, M.},
  journal = {Phys. Rev. B},
  volume = {95},
  issue = {9},
  pages = {094120},
  numpages = {11},
  year = {2017},
  month = {Mar},
  publisher = {American Physical Society},
  doi = {10.1103/PhysRevB.95.094120}
}

@article{backofen2024nonequilibrium,
  title={Nonequilibrium hyperuniform states in active turbulence},
  author={Backofen, Rainer and Altawil, Abdelrahman YA and Salvalaglio, Marco and Voigt, Axel},
  journal={Proc. Natl. Acad. Sci. U.S.A.},
  volume={121},
  number={24},
  pages={e2320719121},
  doi={10.1073/pnas.2320719121},
  year={2024}
}

@article{kim_ultradense_2025,
    author = {Kim, Jaeuk and Torquato, Salvatore},
    title = {Ultradense sphere packings derived from disordered stealthy hyperuniform ground states},
    journal = {J. Chem. Phys.},
    volume = {163},
    number = {2},
    pages = {024902},
    year = {2025},
    month = {07},
    issn = {0021-9606},
    doi = {10.1063/5.0276041}
}

@article{cross_pattern_1993,
  author  = {Cross, M. C. and Hohenberg, P. C.},
  title   = {Pattern formation outside of equilibrium},
  journal = {Rev. Mod. Phys.},
  volume  = {65},
  number  = {3},
  pages   = {851--1112},
  year    = {1993},
  doi     = {10.1103/RevModPhys.65.851}
}

@article{swift_hydrodynamic_1977,
  author  = {Swift, J. and Hohenberg, P. C.},
  title   = {Hydrodynamic fluctuations at the convective instability},
  journal = {Phys. Rev. A},
  volume  = {15},
  number  = {1},
  pages   = {319--328},
  year    = {1977},
  doi     = {10.1103/PhysRevA.15.319}
}

@article{turing_chemical_1952,
  author  = {Turing, A. M.},
  title   = {The Chemical Basis of Morphogenesis},
  journal = {Philos. Trans. R. Soc. Lond., B, Biol. Sci.},
  volume  = {237},
  number  = {641},
  pages   = {37--72},
  year    = {1952},
  doi     = {10.1098/rstb.1952.0012}
}

@article{cahn_free_1958,
  author  = {Cahn, J. W. and Hilliard, J. E.},
  title   = {Free Energy of a Nonuniform System. I. Interfacial Free Energy},
  journal = {J. Chem. Phys.},
  volume  = {28},
  number  = {2},
  pages   = {258--267},
  year    = {1958},
  doi     = {10.1063/1.1744102}
}

@article{cahn_spinodal_1961,
  author  = {Cahn, J. W.},
  title   = {On spinodal decomposition},
  journal = {Acta Metall.},
  volume  = {9},
  number  = {9},
  pages   = {795--801},
  year    = {1961},
  doi     = {10.1016/0001-6160(61)90182-1}
}

@article{hohenberg_theory_1977,
  author  = {Hohenberg, P. C. and Halperin, B. I.},
  title   = {Theory of dynamic critical phenomena},
  journal = {Rev. Mod. Phys.},
  volume  = {49},
  number  = {3},
  pages   = {435--479},
  year    = {1977},
  doi     = {10.1103/RevModPhys.49.435}
}

@article{kim2020multifunctional,
  title={Multifunctional composites for elastic and electromagnetic wave propagation},
  author={Kim, Jaeuk and Torquato, Salvatore},
  journal={Proc. Natl. Acad. Sci. U.S.A.},
  volume={117},
  number={16},
  pages={8764--8774},
  year={2020}
}

@article{ma_random_2017,
  title={Random scalar fields and hyperuniformity},
  author={Ma, Z. and Torquato, S.},
  journal={J. Appl. Phys.},
  volume={121},
  number={24},
  doi={10.1063/1.4989492},
  pages={244904},
  year={2017},
  publisher={AIP Publishing LLC}
}

@article{aaron_fast_2024,
  title = {Fast generation of spectrally shaped disorder},
  author = {Shih, Aaron and Casiulis, Mathias and Martiniani, Stefano},
  journal = {Phys. Rev. E},
  volume = {110},
  issue = {3},
  pages = {034122},
  numpages = {19},
  year = {2024},
  month = {Sep},
  publisher = {American Physical Society},
  doi = {10.1103/PhysRevE.110.034122}
}

@article{uche_constraints_2004,
  title = {Constraints on collective density variables: Two dimensions},
  author = {Uche, Obioma U. and Stillinger, Frank H. and Torquato, Salvatore},
  journal = {Phys. Rev. E},
  volume = {70},
  issue = {4},
  pages = {046122},
  numpages = {9},
  year = {2004},
  month = {Oct},
  doi = {10.1103/PhysRevE.70.046122}
}

@article{salvalaglio_persistent_2024,
	title = {Persistent homology and topological statistics of hyperuniform point clouds},
	volume = {6},
	url = {https://link.aps.org/doi/10.1103/PhysRevResearch.6.023107},
	doi = {10.1103/PhysRevResearch.6.023107},
	number = {2},
	urldate = {2025-09-03},
	journal = {Phys. Rev. Res.},
	publisher = {American Physical Society},
	author = {Salvalaglio, Marco and Skinner, Dominic J. and Dunkel, Jörn and Voigt, Axel},
	month = may,
	year = {2024},
	pages = {023107},
}

@article{elder_modeling_2004,
	title = {Modeling elastic and plastic deformations in nonequilibrium processing using phase field crystals},
	volume = {70},
	url = {https://link.aps.org/doi/10.1103/PhysRevE.70.051605},
	doi = {10.1103/PhysRevE.70.051605},
	number = {5},
	urldate = {2025-09-03},
	journal = {Phys. Rev. E},
	publisher = {American Physical Society},
	author = {Elder, K. R. and Grant, Martin},
	month = nov,
	year = {2004},
	pages = {051605},
}

@article{chen_topological_2021,
	title = {Topological defects, inherent structures, and hyperuniformity},
	volume = {104},
	url = {https://link.aps.org/doi/10.1103/PhysRevB.104.174101},
	doi = {10.1103/PhysRevB.104.174101},
	number = {17},
	urldate = {2025-09-04},
	journal = {Phys. Rev. B},
	publisher = {American Physical Society},
	author = {Chen, Duyu and Zheng, Yu and Jiao, Yang},
	month = nov,
	year = {2021},
	pages = {174101},
}

@article{toth_polymorphism_2010,
	title = {Polymorphism, crystal nucleation and growth in the phase-field crystal model in {2D} and {3D}},
	volume = {22},
	issn = {0953-8984},
	url = {https://doi.org/10.1088/0953-8984/22/36/364101},
	doi = {10.1088/0953-8984/22/36/364101},
	
	number = {36},
	urldate = {2025-09-25},
	journal = {J. Phys.: Condens. Matter},
	author = {Tóth, Gyula I and Tegze, György and Pusztai, Tamás and Tóth, Gergely and Gránásy, László},
	month = aug,
	year = {2010},
	pages = {364101},
}

@article{noauthor_isotropic_nodate,
    author  = {Man, Weining and Florescu, Marian and Williamson, Eric Paul and He, Yingquan and Hashemizad, Seyed Reza and Leung, Brian Y. C. and Liner, Devin Robert and Torquato, Salvatore and Chaikin, Paul M. and Steinhardt, Paul J.},
    title   = {Isotropic band gaps and freeform waveguides observed in hyperuniform disordered photonic solids},
    journal = {Proc. Natl. Acad. Sci. U.S.A.},
    volume  = {110},
    number  = {40},
    pages   = {15886--15891},
    year    = {2013},
    doi     = {10.1073/pnas.1307879110}
}

@article{yu_engineered_2021,
	title = {Engineered disorder in photonics},
	volume = {6},
	copyright = {2020 Springer Nature Limited},
	issn = {2058-8437},
	url = {https://www.nature.com/articles/s41578-020-00263-y},
	doi = {10.1038/s41578-020-00263-y},
	
	number = {3},
	urldate = {2025-09-30},
	journal = {Nat. Rev. Mater.},
	publisher = {Nature Publishing Group},
	author = {Yu, Sunkyu and Qiu, Cheng-Wei and Chong, Yidong and Torquato, Salvatore and Park, Namkyoo},
	month = mar,
	year = {2021},
	pages = {226--243},
}

@article{kim_effect_2018,
	title = {Effect of imperfections on the hyperuniformity of many-body systems},
	volume = {97},
	url = {https://link.aps.org/doi/10.1103/PhysRevB.97.054105},
	doi = {10.1103/PhysRevB.97.054105},
	number = {5},
	urldate = {2025-09-30},
	journal = {Phys. Rev. B},
	publisher = {American Physical Society},
	author = {Kim, Jaeuk and Torquato, Salvatore},
	month = feb,
	year = {2018},
	pages = {054105},
}

@article{Nizam_2021,
doi = {10.1088/1361-648X/abf9b8},
year = {2021},
month = {jun},
publisher = {IOP Publishing},
volume = {33},
number = {30},
pages = {304002},
author = {Nizam, \"u Seleme and Makey, Ghaith and Barbier, Micha\"el and Kahraman, S S\"uleyman and Demir, Esin and Shafigh, Ehsan E and Galioglu, Sezin and Vahabli, Danial and H\"usn\"ugil, Sercan and G\"uneş, Muhammed H and Yelesti, Efe and Ilday, Serim},
title = {Dynamic evolution of hyperuniformity in a driven dissipative colloidal system},
journal = {J. Phys. Condens. Matter},
}

@article{xtal1,
title="Free Energy Functionals for Efficient Phase Field Crystal 
Modeling of Structural Phase Transformations",
author="M. Greenwood and N. Provatas and J. Rottler",
journal="Phys. Rev. Lett.",volume="105",pages=045702,year=2010,doi={10.1103/PhysRevLett.105.045702}}

@article{xtal2,
title="Phase-field-crystal methodology for modeling of structural transformations",
author="M. Greenwood and J. Rottler and N. Provatas",
journal="Phys. Rev. E",volume="83",pages=031601,year=2011,doi={10.1103/PhysRevE.83.031601}}

@article{chen_disordered_2023,
	title = {Disordered hyperuniform solid state materials},
	volume = {10},
	issn = {1931-9401},
	url = {https://doi.org/10.1063/5.0137187},
	doi = {10.1063/5.0137187},
	number = {2},
	urldate = {2025-10-02},
	journal = {Appl. Phys. Rev.},
	author = {Chen, Duyu and Zhuang, Houlong and Chen, Mohan and Huang, Pinshane Y. and Vlcek, Vojtech and Jiao, Yang},
	month = may,
	year = {2023},
	pages = {021310},
}

@article{Gonzalo_HU_2019,
  title = {Hyperuniform vortex patterns at the surface of type-II superconductors},
  author = {Rumi, Gonzalo and Arag{\'o}n S{\'a}nchez, Jazm{\'{\i}}n and El{\'{\i}}as, Federico and Cort{\'e}s Maldonado, Ra{\'u}l and Puig, Joaqu{\'{\i}}n and Cejas Bolecek, N{\'e}stor Ren{\'e} and Nieva, Gladys and Konczykowski, Marcin and Fasano, Yanina and Kolton, Alejandro B.},
  journal = {Phys. Rev. Res.},
  volume = {1},
  issue = {3},
  pages = {033057},
  numpages = {17},
  year = {2019},
  month = {Oct},
  publisher = {American Physical Society},
  doi = {10.1103/PhysRevResearch.1.033057}
}

@article{piechulla2021tailored,
  author  = {Piechulla, Peter M. and Fuhrmann, Bodo and
             Slivina, Evgeniia and Rockstuhl, Carsten and
             Wehrspohn, Ralf B. and Sprafke, Alexander N.},
  title   = {Tailored Light Scattering through Hyperuniform
             Disorder in Self-Organized Arrays of High-Index Nanodisks},
  journal = {Adv. Opt. Mater.},
  volume  = {9},
  number  = {17},
  pages   = {2100186},
  year    = {2021},
  doi     = {10.1002/adom.202100186}
}

@article{yu2021engineered,
  title={Engineered disorder in photonics},
  author={Yu, S. and Qiu, C.-W. and Chong, Y. and Torquato, S. and Park, N.},
  journal={Nat. Rev. Mater.},
  volume={6},
  number={3},
  pages={226--243},
  doi={10.1038/7257780a},
  year={2021},
  publisher={Nature Publishing Group UK London}
}

@article{Vynck2023,
  title = {Light in correlated disordered media},
  author = {Vynck, K. and Pierrat, R. and Carminati, R. and Froufe-P\'erez, L. S. and Scheffold, F. and Sapienza, R. and Vignolini, S. and S\'aenz, J. J.},
  journal = {Rev. Mod. Phys.},
  volume = {95},
  issue = {4},
  pages = {045003},
  numpages = {63},
  year = {2023},
  publisher = {American Physical Society},
  doi = {10.1103/RevModPhys.95.045003}
}

@article{marcotte_nonequilibrium_2013,
	title = {Nonequilibrium static growing length scales in supercooled liquids on approaching the glass transition},
	volume = {138},
	issn = {0021-9606},
	url = {https://doi.org/10.1063/1.4769422},
	doi = {10.1063/1.4769422},
	number = {12},
	urldate = {2025-10-04},
	journal = {J. Chem. Phys.},
	author = {Marcotte, {\'E}tienne and Stillinger, Frank H. and Torquato, Salvatore},
	month = jan,
	year = {2013},
	pages = {12A508},
}

@article{zhang_perfect_2016,
	title = {The {Perfect} {Glass} {Paradigm}: {Disordered} {Hyperuniform} {Glasses} {Down} to {Absolute} {Zero}},
	volume = {6},
	copyright = {2016 The Author(s)},
	issn = {2045-2322},
	shorttitle = {The {Perfect} {Glass} {Paradigm}},
	url = {https://www.nature.com/articles/srep36963},
	doi = {10.1038/srep36963},
	
	number = {1},
	urldate = {2025-10-04},
	journal = {Sci. Rep.},
	publisher = {Nature Publishing Group},
	author = {Zhang, G. and Stillinger, F. H. and Torquato, S.},
	month = nov,
	year = {2016},
	pages = {36963},
}

@article{mkhonta_exploring_2013,
	title = {Exploring the {Complex} {World} of {Two}-{Dimensional} {Ordering} with {Three} {Modes}},
	volume = {111},
	url = {https://link.aps.org/doi/10.1103/PhysRevLett.111.035501},
	doi = {10.1103/PhysRevLett.111.035501},
	number = {3},
	urldate = {2025-10-09},
	journal = {Phys. Rev. Lett.},
	publisher = {American Physical Society},
	author = {Mkhonta, S. K. and Elder, K. R. and Huang, Zhi-Feng},
	month = jul,
	year = {2013},
	pages = {035501},
}

@article{steinhardt_bond-orientational_1983,
	title = {Bond-orientational order in liquids and glasses},
	volume = {28},
	url = {https://link.aps.org/doi/10.1103/PhysRevB.28.784},
	doi = {10.1103/PhysRevB.28.784},
	number = {2},
	urldate = {2025-10-09},
	journal = {Phys. Rev. B},
	publisher = {American Physical Society},
	author = {Steinhardt, Paul J. and Nelson, David R. and Ronchetti, Marco},
	month = jul,
	year = {1983},
	pages = {784--805},
}

@article{zhang_approach_2023,
	title = {Approach to hyperuniformity in a metallic glass-forming material exhibiting a fragile to strong glass transition},
	volume = {46},
	issn = {1292-895X},
	url = {https://doi.org/10.1140/epje/s10189-023-00308-4},
	doi = {10.1140/epje/s10189-023-00308-4},
	
	number = {6},
	urldate = {2025-10-17},
	journal = {Eur. Phys. J. E},
	author = {Zhang, Hao and Wang, Xinyi and Zhang, Jiarui and Yu, Hai-Bin and Douglas, Jack F.},
	month = jun,
	year = {2023},
	pages = {50},
}

@book{hansen_theory_2013,
	title = {Theory of {Simple} {Liquids}: {With} {Applications} to {Soft} {Matter}},
	isbn = {978-0-12-387032-2 978-0-12-387033-9},
	shorttitle = {Theory of {Simple} {Liquids}},
	url = {https://www.sciencedirect.com/book/monograph/9780123870322/theory-of-simple-liquids},
	doi = {10.1016/C2010-0-66723-X},
	urldate = {2026-05-29},
	publisher = {Academic Press},
	author = {Hansen, Jean-Pierre and McDonald, Ian R.},
	month = aug,
	year = {2013},
}

@article{toth_amorphous_2011,
	title = {Amorphous {Nucleation} {Precursor} in {Highly} {Nonequilibrium} {Fluids}},
	volume = {107},
	url = {https://link.aps.org/doi/10.1103/PhysRevLett.107.175702},
	doi = {10.1103/PhysRevLett.107.175702},
	number = {17},
	urldate = {2026-05-29},
	journal = {Phys. Rev. Lett.},
	publisher = {American Physical Society},
	author = {Tóth, Gyula I. and Pusztai, Tamás and Tegze, György and Tóth, Gergely and Gránásy, László},
	month = oct,
	year = {2011},
	pages = {175702},
}

@article{berry_modeling_2011,
	title = {Modeling {Multiple} {Time} {Scales} during {Glass} {Formation} with {Phase}-{Field} {Crystals}},
	volume = {106},
	url = {https://link.aps.org/doi/10.1103/PhysRevLett.106.175702},
	doi = {10.1103/PhysRevLett.106.175702},
	number = {17},
	urldate = {2026-05-29},
	journal = {Phys. Rev. Lett.},
	publisher = {American Physical Society},
	author = {Berry, Joel and Grant, Martin},
	month = apr,
	year = {2011},
	pages = {175702},
}

@article{berry_phase-field-crystal_2014,
	title = {Phase-field-crystal modeling of glass-forming liquids: {Spanning} time scales during vitrification, aging, and deformation},
	volume = {89},
	shorttitle = {Phase-field-crystal modeling of glass-forming liquids},
	url = {https://link.aps.org/doi/10.1103/PhysRevE.89.062303},
	doi = {10.1103/PhysRevE.89.062303},
	number = {6},
	urldate = {2026-05-29},
	journal = {Phys. Rev. E},
	publisher = {American Physical Society},
	author = {Berry, Joel and Grant, Martin},
	month = jun,
	year = {2014},
	pages = {062303},
}

@article{berry_simulation_2008,
	title = {Simulation of an atomistic dynamic field theory for monatomic liquids: {Freezing} and glass formation},
	volume = {77},
	shorttitle = {Simulation of an atomistic dynamic field theory for monatomic liquids},
	url = {https://link.aps.org/doi/10.1103/PhysRevE.77.061506},
	doi = {10.1103/PhysRevE.77.061506},
	number = {6},
	urldate = {2026-05-29},
	journal = {Phys. Rev. E},
	publisher = {American Physical Society},
	author = {Berry, Joel and Elder, K. R. and Grant, Martin},
	month = jun,
	year = {2008},
	pages = {061506},
}

@article{elder_phase-field_2007,
	title = {Phase-field crystal modeling and classical density functional theory of freezing},
	volume = {75},
	url = {https://link.aps.org/doi/10.1103/PhysRevB.75.064107},
	doi = {10.1103/PhysRevB.75.064107},
	number = {6},
	urldate = {2026-05-29},
	journal = {Phys. Rev. B},
	publisher = {American Physical Society},
	author = {Elder, K. R. and Provatas, Nikolas and Berry, Joel and Stefanovic, Peter and Grant, Martin},
	month = feb,
	year = {2007},
	pages = {064107},
}

@article{mkhonta_liquid-substrate_2024,
	title = {Liquid-substrate fluctuation effects on crystal growth and disordered hyperuniformity of two-dimensional materials},
	volume = {8},
	url = {https://link.aps.org/doi/10.1103/PhysRevMaterials.8.104002},
	doi = {10.1103/PhysRevMaterials.8.104002},
	number = {10},
	urldate = {2026-05-29},
	journal = {Phys. Rev. Mater.},
	publisher = {American Physical Society},
	author = {Mkhonta, S. K. and Huang, Zhi-Feng and Elder, K. R.},
	month = oct,
	year = {2024},
	pages = {104002},
}

@article{skogvoll_phase_2022,
	title = {A phase field crystal theory of the kinematics of dislocation lines},
	volume = {166},
	issn = {0022-5096},
	url = {https://www.sciencedirect.com/science/article/pii/S0022509622001296},
	doi = {10.1016/j.jmps.2022.104932},
	urldate = {2026-05-29},
	journal = {J. Mech. Phys. Solids},
	author = {Skogvoll, Vidar and Angheluta, Luiza and Skaugen, Audun and Salvalaglio, Marco and Viñals, Jorge},
	month = sep,
	year = {2022},
	pages = {104932},
}

@article{elder_modeling_2002,
	title = {Modeling {Elasticity} in {Crystal} {Growth}},
	volume = {88},
	url = {https://link.aps.org/doi/10.1103/PhysRevLett.88.245701},
	doi = {10.1103/PhysRevLett.88.245701},
	number = {24},
	urldate = {2026-05-29},
	journal = {Phys. Rev. Lett.},
	publisher = {American Physical Society},
	author = {Elder, K. R. and Katakowski, Mark and Haataja, Mikko and Grant, Martin},
	month = jun,
	year = {2002},
	pages = {245701},
}

@article{torquato_hyperuniform_2018,
	title = {Hyperuniform states of matter},
	volume = {745},
	issn = {0370-1573},
	doi = {10.1016/j.physrep.2018.03.001},
	journal = {Phys. Rep.},
	author = {Torquato, Salvatore},
	month = jun,
	year = {2018},
	pages = {1--95}
}

@article{abdalla_thermodynamics_2022,
	title = {Thermodynamics, formation dynamics, and structural correlations in the bulk amorphous phase of the phase-field crystal model},
	volume = {157},
	issn = {0021-9606},
	url = {https://doi.org/10.1063/5.0114705},
	doi = {10.1063/5.0114705},
	number = {16},
	urldate = {2026-05-11},
	journal = {J. Chem. Phys.},
	author = {Abdalla, Shaho and Archer, Andrew J. and Gránásy, László and Tóth, Gyula I.},
	month = oct,
	year = {2022},
	pages = {164502},
}

@article{galenko_marginal_2011,
	title = {Marginal stability analysis of the phase field crystal model in one spatial dimension},
	volume = {83},
	url = {https://link.aps.org/doi/10.1103/PhysRevB.83.064113},
	doi = {10.1103/PhysRevB.83.064113},
	number = {6},
	urldate = {2026-08-31},
	journal = {Phys. Rev. B},
	publisher = {American Physical Society},
	author = {Galenko, P. K. and Elder, K. R.},
	month = feb,
	year = {2011},
	pages = {064113},
}

@article{berthier_theoretical_2011,
	title = {Theoretical perspective on the glass transition and amorphous materials},
	volume = {83},
	url = {https://link.aps.org/doi/10.1103/RevModPhys.83.587},
	doi = {10.1103/RevModPhys.83.587},
	number = {2},
	urldate = {2026-06-12},
	journal = {Rev. Mod. Phys.},
	publisher = {American Physical Society},
	author = {Berthier, Ludovic and Biroli, Giulio},
	month = jun,
	year = {2011},
	pages = {587--645},
}

@article{debenedetti_supercooled_2001,
	title = {Supercooled liquids and the glass transition},
	volume = {410},
	copyright = {2001 Springer Nature Limited},
	issn = {1476-4687},
	url = {https://www.nature.com/articles/35065704},
	doi = {10.1038/35065704},
	
	number = {6825},
	urldate = {2026-09-13},
	journal = {Nature},
	publisher = {Nature Publishing Group},
	author = {Debenedetti, Pablo G. and Stillinger, Frank H.},
	month = mar,
	year = {2001},
	pages = {259--267},
}

@article{torquato_local_2003,
	title = {Local density fluctuations, hyperuniformity, and order metrics},
	volume = {68},
	url = {https://link.aps.org/doi/10.1103/PhysRevE.68.041113},
	doi = {10.1103/PhysRevE.68.041113},
	number = {4},
	urldate = {2026-09-13},
	journal = {Phys. Rev. E},
	publisher = {American Physical Society},
	author = {Torquato, Salvatore and Stillinger, Frank H.},
	month = oct,
	year = {2003},
	pages = {041113},
}

@article{uche_collective_2006,
	title = {Collective coordinate control of density distributions},
	volume = {74},
	url = {https://link.aps.org/doi/10.1103/PhysRevE.74.031104},
	doi = {10.1103/PhysRevE.74.031104},
	number = {3},
	urldate = {2026-09-17},
	journal = {Phys. Rev. E},
	publisher = {American Physical Society},
	author = {Uche, Obioma U. and Torquato, Salvatore and Stillinger, Frank H.},
	month = sep,
	year = {2006},
	pages = {031104},
}

\end{document}